\documentclass[showpacs,apl,reprint,amsmath,amssymb,superscriptaddress,nobibnotes,twocolumn, preprintnumbers, bibnotes]{revtex4-1}
\usepackage{epsfig}
\usepackage{graphicx}
\usepackage{amsmath}
\usepackage{bbold}
\usepackage{tabularx, booktabs}
\usepackage[normalem]{ulem}
\usepackage{xcolor}
\usepackage{multirow}
\usepackage{dutchcal} 
\usepackage[T1]{fontenc}
\usepackage{comment}
\usepackage{longtable}
\usepackage{appendix}
\usepackage[colorlinks=true,allcolors=black]{hyperref}
\usepackage{chemformula}
\let\ce\ch
\usepackage{orcidlink}

\usepackage{xr}
\usepackage{chemformula}
\let\ce\ch

\usepackage{float}
\usepackage[caption = false]{subfig}
\newcolumntype{Y}{>{\centering\arraybackslash}X}
\newcolumntype{Z}{>{\hsize=1.1\hsize\centering\arraybackslash}X}
\newcommand{\Neel}{N\'{e}el}

\begin{document}
\newcommand{\xidian}{School of Advanced Materials and Nanotechnology, XiDian University, Xi’an 710126, China}
\newcommand{\xihang}{School of Electronic Engineering, Xi’an Aeronautical Institute, Xi’an 710077, China}
\newcommand{\nwpu}{School of Materials Science and Engineering, Northwestern Polytechnical University, Xi’an 710072, China}
\newcommand{\upv}{Fisika Aplikatua Saila, Gipuzkoako Ingeniaritza Eskola, University of the Basque Country (UPV/EHU), 
		Europa Plaza 1, 20018 Donostia/San Sebasti{\'a}n, Spain}
\newcommand{\cfm}{Centro de F{\'i}sica de Materiales (CFM-MPC), CSIC-UPV/EHU,  Manuel de Lardizabal Pasealekua 5, 20018 Donostia/San Sebasti{\'a}n, Spain}

\title{
Control of magnetic transition, metal-semiconductor transition, and magnetic anisotropy in noncentrosymmetric monolayer  Cr$_2$Ge$_2$Se$_3$Te$_3$
}

\author{Rui-Qi Wang\orcidlink{0000-0003-4725-3892}}
\affiliation{\xidian}
\affiliation{\xihang}
\author{Tengfei Cao\orcidlink{0000-0001-9508-2966}}
\affiliation{\nwpu}
\author{Tian-Min Lei\orcidlink{0000-0001-6377-6218}}
\affiliation{\xidian}
\author{Xie Zhang\orcidlink{0000-0003-3424-9253}}
\email{xie.zhang@nwpu.edu.cn}
\affiliation{\nwpu}
\author{Yue-Wen Fang\orcidlink{0000-0003-3674-7352}}
\email{yuewen.fang@csic.es or yuewen.fang@ehu.eus} 
\affiliation{\cfm}

\begin{abstract}
Recent advances in two-dimensional materials have greatly expanded the family of ferromagnetic materials.  The well-known 2D ferromagnets, such as \ce{CrI3}, \ce{Cr2Ge2Te6}, and \ce{Fe3GeTe2} monolayers, are characterized by centrosymmetric crystal structures. In contrast, ferromagnetic ordering in 2D noncentrosymmetric materials remains an underexplored area. Here we report a Janus ferromagnet, \ce{Cr2Ge2Se3Te3} with inversion symmetry breaking, through first-principles calculations. This monolayer can undergo a ferromagnetic-antiferromagnetic transformation and a metal-semiconductor transition under different strains. Additionally, the strength of magnetocrystalline anisotropy energy (MAE) can be modulated by electric field or strain. In particular, the magnetization easy axis can be altered from in-plane to out-of-plane under strain. We find that Te$_3$ atoms play a key role in determining the MAE, where contributions are primarily from  $p_z / p_y$ and $p_x / p_y$ orbitals. This study of Janus ferromagnetic materials has provided a promising platform for the research on the control of magnetism by strain or electric field.
\end{abstract}
\maketitle
\newpage

To date,  many well-known 2D ferromagnets, such as CrI$_3$, Cr$_2$Ge$_2$Te$_6$, VSe$_2$, and Fe$_3$GeTe$_2$ monolayers have been extensively studied~\cite{CrI3-nature-2017,CGT-nature-2017,VSe2-Nature-2018,FGT-Nature-Material-2018,Xin-Chip-2D-ferro2023}. However, these materials typically have low magnetic transition temperatures or weak magnetic anisotropy (MA), which limit their applications. In order to improve these properties, various methods have been employed, such as applying an electric field\cite{Niu-AM-2024,Bednarz-2024-efiled-APL}, using ferroelectric polarization\cite{FE-NanoLett-2024,Zhang-FE-APL,wang-JAP-2024}, applying strain\cite{strain-PRB-2024,Li-strain-2024,An-JMMM-2024}, and doping\cite{Hu-dope-PRB-2024,dope-PRB-2024}. These methods are also possible to drive transitions from antiferromagnetic (AFM) ordering to ferromagnetic (FM) ordering\cite{Zhou-Nanoletter-2023,Qiao-PRB-2023,Wang-PRA-2024,Liu-ACS-Applied-Electronic-Materials-2023}, as well as the reorientation of the magnetization easy axis by 90$^{\circ}$\cite{Wang-NPJ-2022,JAP-2024-Co-CrI3,PCCP-2024-WSe2,Liu-Nanoscale-2022,Jin-app-sur-sci-2022,Yao-APL-2023}.\\
\indent 
In the various methods, the substitution of non-magnetic atoms of 2D magnets has become a key approach to make them Janus magnets, in which mirror symmetry is broken by having different atomic species on the top and bottom layers\cite{Vojacek-2024-Nanoletter,Wan-PCCP-2023,Liu-PCCP-2023,Zhang-2D-2023}. For instance, by substituting the I atoms with different halogen atoms in \ce{CrI3}, Cr${XY_2}$ ($XY$= IBr, ICl, BrCl) can be obtained\cite{Chen-pss-2022}. Replacing the Se atoms with other chalcogen atoms in \ce{VSe2} yields VSSe, VSTe or VSeTe\cite{Wang-MN-2023,Guan-Nanoscale-2020}. Similarly, substituting the Te atoms in \ce{CrTe2} yields CrSSe, CrSTe, or CrSeTe\cite{Cui-PRB-2020,Vojacek-2024-Nanoletter}. In addition, \ce{Fe3GeSeTe} can be obtained by substituting one of the Te atoms in \ce{Fe3GeTe2} with Se\cite{Chen-APL-2023}. Furthermore, the Janus ferromagnet \ce{Cr2}$XY$\ce{Te6} ($X$, $Y$ = Si, Ge, Sn, and $X \neq Y$) can been obtained by setting $X$ and $Y$ to be different IVA elements\cite{Xiao-JAP-2024}. Among these Janus ferromagnets, FM-AFM transitions\cite{Liu-APL-2023, Xiao-JAP-2024, Rahman-JEM-2023, Esteras-Nano-2022} 
or reorientation of the magnetization easy axis by 90$^{\circ}$~\cite{Liu-APL-2023,Yao-APL-2023,Wang-MN-2023,Rudenko-njp-2023} are achieved through strain engineering. In addition, the noncentrosymmetric structure is conducive to Dzyaloshinskii-Moriya interaction (DMI) effect\cite{Cui-PRB-2020}, Rashba-type or beyond-Rashba-type large spin splittings\cite{Lee-PRB-2024, PRB-splitting-LWN2020}, or multi-fields control\cite{Liu-APL-2023}. \\
\indent 
Herein, we predict a Janus ferromagnet \ce{Cr2Ge2Se3Te3} by replacing the Te atoms with Se atoms on one side of the symmetric ferromagnet \ce{Cr2Ge2Te6}. We investigate the stability of the \ce{Cr2Ge2Se3Te3} and achieve strain-controlled FM-AFM transition. In addition, the manipulation of magnetocrystalline anisotropy energy (MAE) in the Janus ferromagnet is also realized by strain or electric field.

First-principles calculations based on density functional theory (DFT) are performed by using the Vienna ab initio Simulation Package (VASP)~\cite{Kresse1996,Kresse-PRB-1996}. 
The generalized gradient approximation (GGA) of the Perdew–Burke–Ernzerhof (PBE) form is used with an effective Hubbard $U$ = 2.0 eV applied to 3\textit{d} electrons of the Cr atoms\cite{Sui-PRB-2017,Solovyev-PRB-1994}. The convergence criteria for forces and energies during structural relaxations are 10$^{-3}$ eV/\AA \ and 10$^{-8}$ eV, respectively. A 12 $\times$ 12 $\times$ 1 Monkhorst-Pack \textbf{k}-point grids and a 480 eV cut-off energy are used to provide good convergence in self-consistent field calculations. A vacuum layer with a thickness of 25 \AA \ is employed to build the 2D system. The optimized lattice constant for the \ce{Cr2Ge2Se3Te3} monolayer is 6.634 \AA. The Wyckoff positions of the \ce{Cr2Ge2Se3Te3} monolayer is added to the supplementary material as Tab. S1. The Curie (\Neel) temperature is predicted by Monte Carlo simulations using the MCSOLVER code~\cite{Liu-ASS-2019}, and the MAE is determined by the total energy difference between two different spin directions, i.e., MAE = $E$[100] $-$ $E$[001]. 

\begin{figure}
    \centering
    \includegraphics[width=0.99\linewidth]{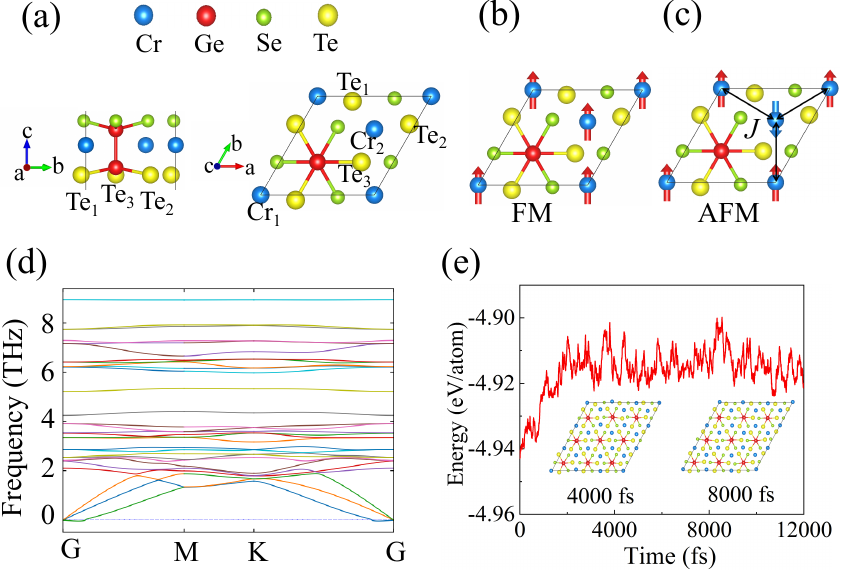}
    \caption{(a) Atomic structures of the \ce{Cr2Ge2Se3Te3} monolayer. The Cr, Ge, Se, and Te atoms are represented by blue, red, green, and yellow spheres, respectively. There are three Te atoms in the primitive cell, labeled as Te$_1$, Te$_2$ and Te$_3$. (b) FM (c) AFM orderings of \ce{Cr2Ge2Se3Te3}. The exchange coupling parameter $J$ is indicated by black arrows. (d) Phonon spectrum of the \ce{Cr2Ge2Se3Te3} monolayer. (e) Variation of the energy from 0 to 12000 fs during MD simulations at 300 K for the \ce{Cr2Ge2Se3Te3} monolayer. The inset illustrates the structural snapshots of the MD simulation at 4000 and 8000 fs, respectively.}
    \label{fig:APL-Figure1}
\end{figure}

The \ce{Cr2Ge2Se3Te3} monolayer has a noncentrosymmetric space group \textit{P}31\textit{m}, and its geometric structure of \ce{Cr2Ge2Se3Te3} is illustrated in Fig. \ref{fig:APL-Figure1}(a). 
By performing the pseudosymmetry search~\cite{Capillas2011-PSEUDO}, we find that another noncentrosymmetric phase with space group of \textit{P}6\textit{mm} is the minimal supergroup of the \textit{P}31\textit{m} phase, indicating potential structure transition between the two structures. The crystal structure of \textit{P}6\textit{mm} is shown in Fig. S1(a) in the supplementary material. As expected, Fig. S1(b) shows that the variation of energy versus the normalized displacement between the low-symmetry \textit{P}31\textit{m} and high-symmetry \textit{P}6\textit{mm} is continuous and smooth. However, this energy barrier of 450 meV per atom is very large, suggesting that the structure transition is unlikely to happen near the room temperature based on the transition state theory~\cite{Fang-Chen-2020BPTO}. Even if we use strain to tune the two phases, we find from Fig. S1(d) that the structure transition is only possible at elevated temperature.
Therefore, we will focus our study on the \textit{P}31\textit{m} structure only.
We consider ferromagnetic ordering (FM) and antiferromagnetic ordering (AFM) of the \textit{P}31\textit{m} structure in Figs. \ref{fig:APL-Figure1}(b) and \ref{fig:APL-Figure1}(c) to find the most favorable magnetic state. The reason why we choose this AFM structure is mainly that there is only one AFM configuration in the current unit cell. We have also conducted Monte Carlo simulations and found that the magnetic configuration maintains the FM state under 0\% strain. The total energy of the FM state is lower than that of AFM state by 58 meV, suggesting the ground state of \ce{Cr2Ge2Se3Te3} monolayer is the FM state.\\
\indent 
To investigate the dynamics stability and the thermodynamic stability, the lattice dynamics calculations and the molecular dynamics (MD) simulation of the FM state are performed, and the results are shown in Figs. \ref{fig:APL-Figure1}(d) and \ref{fig:APL-Figure1}(e). Although there are spoon-shaped curves near the Gamma point [see Fig. \ref{fig:APL-Figure1}(d)], they do not mean instability. Similar soft modes 
are found in other 2D sheets, which are caused by the difficulty converging in 2D materials\cite{Zhao-2017-JPCC,Yi-ASS-2021,Mohebpour-2020-sci-rep}. In addition, we perform MD simulations with a 3 $\times$ 3 $\times$ 1 supercell in two steps. In the initial step, we simulate the process of gradual temperature increase from 0 to 300 K. Next, we perform MD simulations at 300 K using the structure from the former step. As a result, the energy changes are small (from $-4.92$ to $-4.90$ eV/atom) and exhibit periodic vibrations [see Fig. \ref{fig:APL-Figure1}(e)]. Further, we used the equation (\ref{eq:H}) to calculate the formation enthalpy, where the energy of each element has been normalized to per atom. It is calculated to be -0.453 eV/atom, which proves that the \ce{Cr2Ge2Se3Te3} is likely to form and remains stable rather than decomposing back into its elements.

\begin{equation}
\begin{aligned}
\Delta H = & [ E_{\rm total} - (2E_{\rm bulk} (\mathrm{Cr}) + \\
& 2E_{\rm bulk}(\mathrm{Ge}) + 3E_{\rm bulk}(\mathrm{Se}) + 3E_{\rm bulk}(\mathrm{Te}))] /10
\end{aligned}
\label{eq:H}
\end{equation}

\begin{figure}
    \centering
    \includegraphics[width=0.99\linewidth]{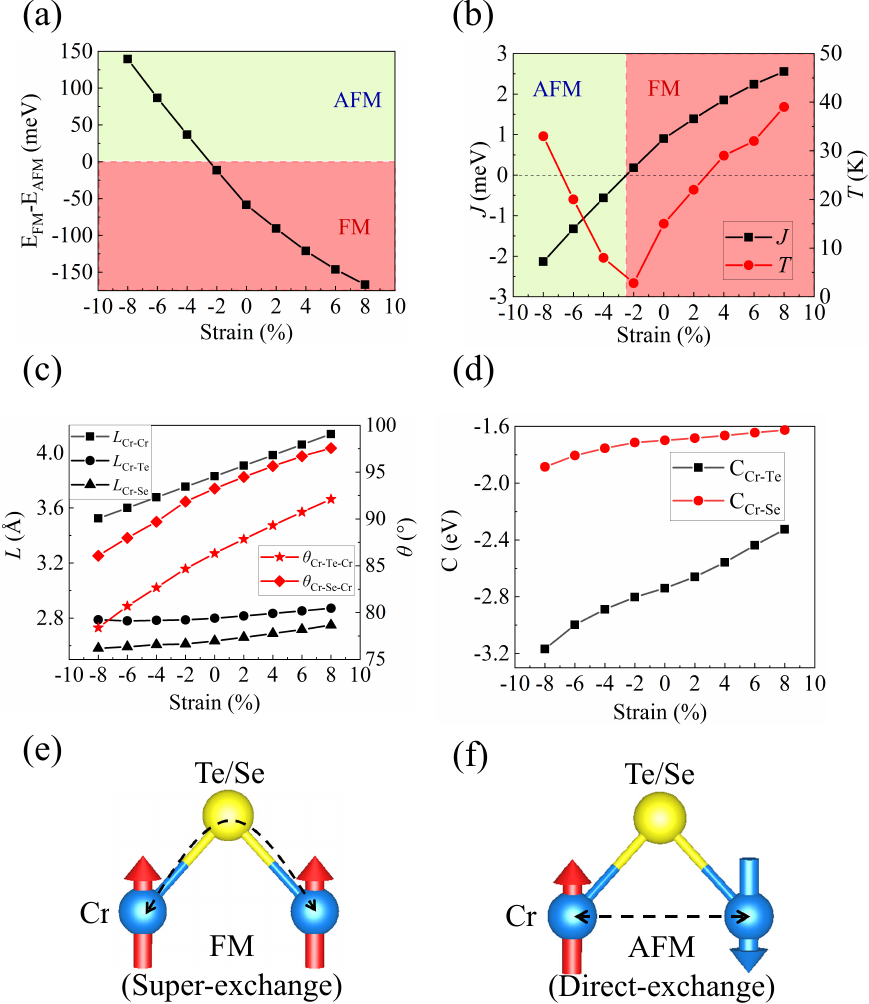}
    \caption{(a) The energy difference between FM state and AFM ground state as a function of the strain. (b) The exchange parameter and transition temperature as functions of the strain. (c) The $L_{\text{Cr-Cr}}$, $L_{\text{Cr-Te/Se}}$, $\theta_{\text{Cr-Te-Cr}}$ and $\theta_{\text{Cr-Se-Cr}}$ as functions of the strain. (d) The covalency between Cr and Te/Se atoms as a function of the strain. (e) The schematic diagrams of the FM Cr–Te/Se–Cr super-exchange. (f) The schematic diagrams of the AFM Cr–Cr direct-exchange interactions.
}
    \label{fig:APL-Figure2}
\end{figure}

We now investigate the impact of biaxial strain on magnetic ordering. Fig. \ref{fig:APL-Figure2}(a) shows the energy differences (unit cell) between the FM and AFM states as a function of the applied strain. It is evident that a compressive strain of over $-2\%$ results in a transition from the FM state to the AFM state.\\
\indent 
The Curie temperature for ferromagnetic materials and \Neel~temperature for antiferromagnetic materials are crucial indicators in understanding how spontaneous magnetization is lost as these materials transition to a paramagnetic state at high temperatures.
In this study, these temperatures are collectively referred to as the phase transition temperatures ($T$) and are solved using Monte Carlo simulations.\\
\indent 
It is known that the Hamiltonian can be expressed as \( H=-J\sum\limits_{ij}S_i S_j\ - A\sum\limits_{i}(S_i^z)^2  - \sum\limits_{ij} D_{ij} \cdot (S_i \times S_j)\) from the Heisenberg model, where $S_i$ and $S_j$ are the spin vectors of the $i$th Cr atom and $j$th Cr atom, respectively. $J$, $A$ and $D_{ij}$  represent the nearest-neighbor exchange coupling parameter, single-ion anisotropy, and Dzyaloshinskii-Moriya interaction, respectively.
Consequently, the total energies of \ce{Cr2Ge2Se3Te3} with FM and AFM orderings can be expressed as:
\begin{equation}
H_{\rm FM}=-3 J S^2-A S^2+E_{\rm other}
\end{equation}
\begin{equation}
H_{\rm AFM}=3 J S^2-A S^2+E_{\rm other}
\end{equation}
Where \textit{H}$_{\rm FM}$ and \textit{H}$_{\rm AFM}$ are the total energy of the FM and AFM configurations, respectively. \textit{E}$_{other}$ includes the DMI term  and the total energy without magnetic coupling. Therefore, \(J = -(H_{\rm FM}-H_{\rm AFM})/6S^2\) [see Fig. \ref{fig:APL-Figure2}(b)]. $A$ and $D_{ij}$ are obtained using
the four-state energy-mapping analysis\cite{Xianghongjun-2013} [see Fig. S2 in the supplementary material]. By performing Monte Carlo simulations, the dependence of the transition temperature on strain is plotted [see Fig. \ref{fig:APL-Figure2}(b)]. We adopt a 20 $\times$ 20 $\times$ 1 supercell in the Monte Carlo calculations with the number of sweeps set to 80,000. It is evident that the transition temperature obtained by 
the Heisenberg model
dose not exceed 40 K. In addition, the transition temperature reaches its minimum at a compressive strain around $-2\%$, which is also the critical point at which the magnetic ordering changes. In the Supplementary Material with a 2 $\times$ 2 $\times$ 1 cell, we artificially choose two classic low-energy AFM configurations, i.e., Néel and Zigzag in Fig. S3(b) and S3(c). With these AFM configurations and the corresponding FM structure, it is sufficient to obtain both the nearest neighbor coupling $J_1$ and the next-nearest neighbor coupling $J_2$ values through solving the Hamiltonian equations. Since the next-nearest neighbor $J_2$ does not significantly influence the transition temperature [see Fig. S4(c)], we only consider the nearest neighbor $J_1$ here and mark it as $J$.

As seen in Fig. \ref{fig:APL-Figure2}(b), the in-plane strain can effectively tune the symbol and strength of the magnetic exchange $J$. In particular, a compressive strain of around $-2\%$ leads to the reversal of $J$, which makes the FM state more favorable than the AFM state, as established in Fig. \ref{fig:APL-Figure2}(a).
Regardless of whether the magnetic coupling is ferromagnetic or antiferromagnetic one, we find that the increase in the magnitude of $J$ leads to the increased transition temperature $T$ at the studied strain range.

Figs. \ref{fig:APL-Figure2}(c)-\ref{fig:APL-Figure2}(d) show the lengths of the Cr-Cr and Cr-Te/Se bonds ($L_{\text{Cr-Cr}}$ and $L_{\text{Cr-Te/Se}}$),  the Cr-Te/Se-Cr bond angles ($\theta_{\text{Cr-Te-Cr}}$ and $\theta_{\text{Cr-Se-Cr}}$), and the covalency between Cr and Te/Se atoms (C$_{\text{Cr-Te/Se}}$) in the \ce{Cr2Ge2Se3Te3} monolayer with different strains. To evaluate the super exchange strength between Cr and Te/Se atoms quantitatively, we define the covalency as\cite{Hou-MOLECULES-2023}: 
\begin{equation}
{\rm C}_{\text{Cr-Te}}=-|\rm{BC}^{Cr} - \rm{BC}^{Te}|
\end{equation}
\begin{equation}
{\rm C}_{\text{Cr-Se}}=-|\rm{BC}^{Cr} - \rm{BC}^{Se}|
\end{equation}
where BC represents band center: 
\begin{equation}
\rm{BC}=\frac{ \int_{\varepsilon_0}^{\varepsilon_1}\varepsilon p(\varepsilon)d\varepsilon}{ \int_{\varepsilon_0}^{\varepsilon_1}p(\varepsilon)d\varepsilon}
\end{equation}
with the projected density of states (PDOS) \(\textit{p}(\varepsilon)\) and band energy \(\varepsilon\).\\
\indent  
As is revealed in Fig. \ref{fig:APL-Figure2}(c),
with increasing compressive strain, the decrease in the Cr–Cr bond length (L$_{\rm Cr-Cr}$) is much greater than that of the Cr–Te and Cr–Se bond length, strengthening the Cr–Cr direct-exchange AFM coupling [Fig. \ref{fig:APL-Figure2}(f)]. This enhancement in direct-exchange coupling can also been confirmed by the evolution of bond angles shown in Fig. \ref{fig:APL-Figure2}(c), in which both $\theta_{\text{Cr-Te-Cr}}$ and $\theta_{\text{Cr-Se-Cr}}$ decrease with the improved compression. In contrast, with increasing tensile strain, the increase of C$_{\text{Cr-Te}}$ is much greater than that of C$_{\text{Cr-Se}}$ [Fig. \ref{fig:APL-Figure2}(d)], revealing the stronger Cr-Te-Cr FM super-exchange interaction [Fig. \ref{fig:APL-Figure2}(e)].

\begin{figure}
    \centering
    \includegraphics[width=0.99\linewidth]{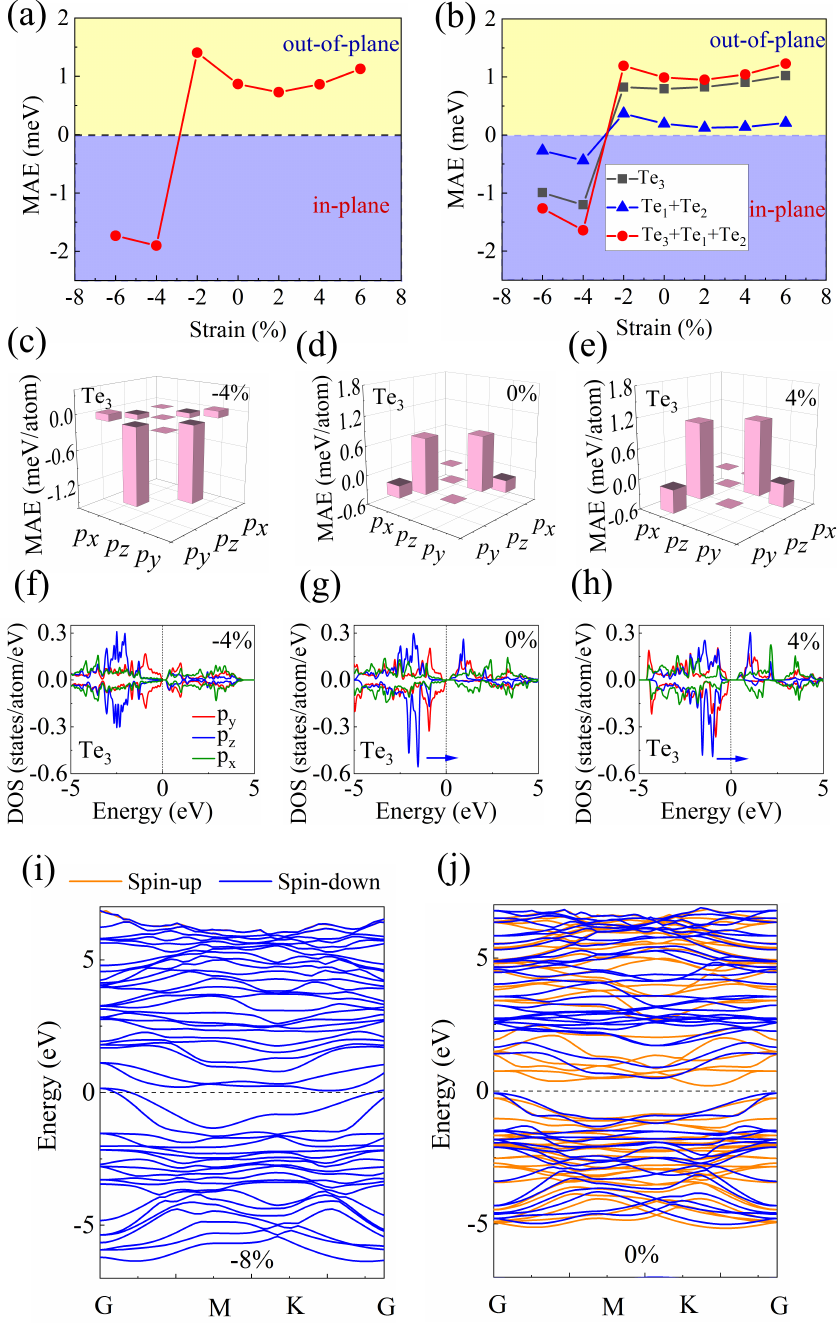}
    \caption{(a) Total MAE dependency on the strain. (b) The atom-resolved MAE dependency on the strain. The \textit{p} orbital-decomposed MAE and PDOS of \ce{Te3} atom in \ce{Cr2Ge2Se3Te3} monolayer under  (c) and (f) $-4\%$ strain, (d) and (g) 0\% strain, and  (e) and (h) 4\% strain. The spin-polarized band structures of \ce{Cr2Ge2Se3Te3} monolayer under (i) $-8\%$ strain and (j) $0\%$ strain.}
    \label{fig:APL-Figure3}
\end{figure}

Next, the total MAE curve with strain is given in Fig. \ref{fig:APL-Figure3}(a), and the contribution of each Te atom is shown in Fig. \ref{fig:APL-Figure3}(b). The MAE will shift from positive to negative when the compressive strain exceeds $-2\%$, as illustrated in Fig. \ref{fig:APL-Figure3}(a). This is also the critical point at which the magnetic ordering shifts from the FM to AFM state. By analyzing the contribution of each atom to the MAE, it can be found that \ce{Te3} contributes the most to the total MAE,  whose MAE is greater than that of the \ce{Te1} and \ce{Te2} atoms [see Fig. \ref{fig:APL-Figure3}(b)]. Here \ce{Te1} and \ce{Te2} atoms are equivalent in their contribution to the total MAE. It should be noted that the total MAE of the system [red curve in Fig. \ref{fig:APL-Figure3}(a)] and the MAE sum of the three Te atoms [red curve in Fig. \ref{fig:APL-Figure3}(b)] are not exactly equal, because there are also minor contributions from Cr, Se and other atoms.\\
\indent  
To demonstrate the contributions of specific orbitals to the MAE, the \textit{p} orbital-decomposed MAE and PDOS for the \ce{Te3} atoms are shown in Figs. \ref{fig:APL-Figure3}(c)-\ref{fig:APL-Figure3}(h). It is clear that the MAE is primarily influenced by two matrices $p_z / p_y$ and $p_x / p_y$. As the strain shifts from $-4\%$ to 0 and then to 4\%, the contribution of $p_z / p_y$ in these matrices changes from negative to positive and experiences a significant increase [see Figs. \ref{fig:APL-Figure3}(c)-\ref{fig:APL-Figure3}(e)]. This is consistent with the total MAE trend of the system, which has shifted from negative value to positive one.\\
\indent
The second-order perturbation theory\cite{Wang-PRB-1993} can be employed to understand the influence of orbitals on the MAE, according to which the MAE can be expressed as:

\begin{equation}
    \text{MAE} = \xi^2 \sum_{o^{+/-}, u^{+/-}} \frac{\left| \langle o^{+/-} | l_z | u^{+/-} \rangle \right|^2 - \left| \langle o^{+/-} | l_x | u^{+/-} \rangle \right|^2}{\epsilon_u^{+/-} - \epsilon_o^{+/-}}
\end{equation}
\begin{equation}
    \text{MAE} = -\xi^2 \sum_{o^{-/+}, u^{+/-}} \frac{\left| \langle o^{-/+} | l_z | u^{+/-} \rangle \right|^2 - \left| \langle o^{-/+} | l_x | u^{+/-} \rangle \right|^2}{\epsilon_u^{+/-} - \epsilon_o^{-/+}}
\end{equation}
where $\xi$ is the spin-orbit coupling (SOC) constant, \textit{o} (\textit{u}) represents the occupied (unoccupied) state, and $\varepsilon_o\left(\varepsilon_u\right)$ denotes the energy level of \textit{o} (\textit{u}) state.
The superscript + and - stand for spin-up and
spin-down channels, respectively. Equation (7) shows the spin-conservation terms ($o^{+}$ $\rightarrow$ $u^{+}$ and $o^{-}$ $\rightarrow$ $u^{-}$) while equation (8) shows the spin-flip terms ($o^{-}$ $\rightarrow$ $u^{+}$ and $o^{+}$ $\rightarrow$ $u^{-}$).

Table \ref{fig:APL-table1} shows the differences of matrix between magnetization along [001] and [100]. By checking all the orbitals in $o^{+}$, $o^{-}$, $u^{+}$, and $u^{-}$ one by one in the PDOS from strain = $-4\%$ to strain = 0\% [see Figs. \ref{fig:APL-Figure3}(f) and \ref{fig:APL-Figure3}(g)], we find that only $o^{-}$ and $u^{+}$ experience obvious change,  where the quantities of $p_z^{o-}$, $p_y^{o-}$, and $p_z^{u+}$ increase, and $p_z^{o-}$ experiences an obvious right shift. Consequently, the MAE variation mainly depends on the $o^{-}$ $\rightarrow$ $u^{+}$ term. Here, the difference of matrix for the $p_z^{o-}$/$p_y^{u+}$ ($p_y^{o-}$/$p_z^{u+}$) hybridization is 1, and that of the $p_z^{o-}$/$p_z^{u+}$ ($p_y^{o-}$/$p_y^{u+}$) hybridization is 0.

As to the MAE under strains from $-4\%$ to 0\%, it increases dramatically from negative to positive.
Firstly, the quantity of $p_z^{o-}$ states increases. Secondly, since the $p_z^{o-}$ states shift to the right [see the blue arrow in Figs. \ref{fig:APL-Figure3}(g)] while the $p_y^{u+}$ states stand still, the value of  $\varepsilon_u-\varepsilon_o$ decreases. Consequently, the MAE increases from the $p_z^{o-}$/$p_y^{u+}$ hybridization since the difference of matrix for it is 1. Furthermore, the quantities of $p_y^{o-}$ and $p_z^{u+}$ states all increase, which yields another increase of MAE from the $p_y^{o-}$/$p_z^{u+}$ hybridization since the difference of matrix for it is 1.

When the strain turns from 0\% to 4\%, the quantities of $p_z^{o-}$, $p_y^{o-}$, and $p_z^{u+}$ generally do not change, whereas the $p_z^{o-}$ states continue to shift to the right [see the blue arrow in Figs. \ref{fig:APL-Figure3}(h)]. As a result, the MAE increases slightly according to the $p_z^{o-}$/$p_y^{u+}$ hybridization.

\begin{table}[htb]
    \caption{The differences of matrix between magnetization along [001] and [100]. The top block ($u^+$) corresponds to $\left| \langle o^+ | l_z | u^+ \rangle \right|^2 - \left| \langle o^+ | l_x | u^+ \rangle \right|^2$ and $\left| \langle o^- | l_x | u^+ \rangle \right|^2 - \left| \langle o^- | l_z | u^+ \rangle \right|^2$. The bottom block ($u^-$) corresponds to $\left| \langle o^- | l_z | u^- \rangle \right|^2 - \left| \langle o^- | l_x | u^- \rangle \right|^2$ and $\left| \langle o^+ | l_x | u^- \rangle \right|^2 - \left| \langle o^+ | l_z | u^- \rangle \right|^2$.}
    \label{fig:APL-table1}
    \centering
    \begin{tabular}{ccrrrrrr}
        \toprule
        \hline
        \multirow{2}{*}{$u$} & \multirow{2}{*}{orbital} & 
        \multicolumn{3}{c}{$o^+$} & \multicolumn{3}{c}{$o^-$} \\
        \cmidrule(lr){3-5} \cmidrule(lr){6-8}
        \cline{3-8}
        & & $p_y$ & $p_z$ & $p_x$ & $p_y$ & $p_z$ & $p_x$ \\
        \hline
        \multirow{3}{*}{$u^+$} 
        & $p_y$ & 0 & $-1$ & 1 & 0 & 1 & $-1$ \\
        & $p_z$ & $-1$ & 0 & 0 & 1 & 0 & 0 \\
        & $p_x$ & 1 & 0 & 0 & $-1$ & 0 & 0 \\
        \midrule
        \multirow{3}{*}{$u^-$} 
        & $p_y$ & 0 & 1 & $-1$ & 0 & $-1$ & 1 \\
        & $p_z$ & 1 & 0 & 0 & $-1$ & 0 & 0 \\
        & $p_x$ & $-1$ & 0 & 0 & 1 & 0 & 0 \\
        \bottomrule
        \hline
    \end{tabular}
\end{table}

In addition, Figs. \ref{fig:APL-Figure3}(f)-\ref{fig:APL-Figure3}(h) reveal that the tensile strain increases the band gap, while the compressive strain decreases it. A critical turning point occurs at a $-7.2\%$ compressive strain where the \ce{Cr2Ge2Se3Te3} undergoes a transition from semiconductors to metals. The band structures of the \ce{Cr2Ge2Se3Te3} monolayer under $-8\%$ strain and $0\%$ strain are shown in Figs. \ref{fig:APL-Figure3}(i) and \ref{fig:APL-Figure3}(j) to demonstrate the metallic and semiconducting states, respectively. Since the stable state of the monolayer under $-8\%$ strain is the AFM state, the spin-up and spin-down states coincide at this point. To study the role of Hund's coupling $J$, we also performed rotationally invariant DFT+$U$ calculations using the approach introduced by Liechtenstein et al~\cite{Liechtenstein1995}. The results are shown in Fig. S5 in the supplementary material. We find that the band gap of the strain-free state, and the critical strain are nearly consistent with the above effective $U$ method, implying our predictions are very robust.

\begin{figure}
    \centering
    \includegraphics[width=0.99\linewidth]{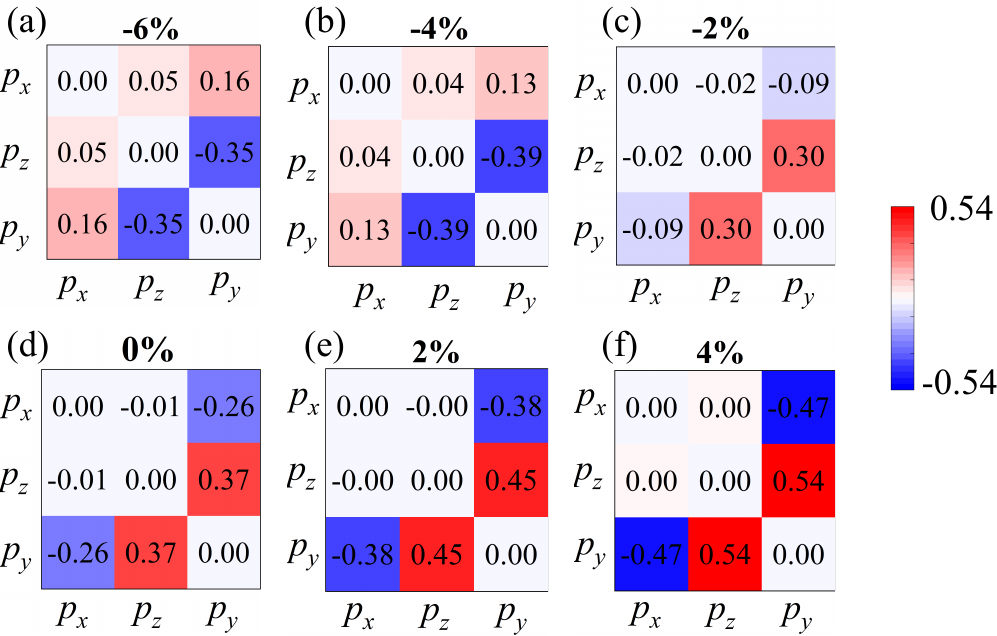}
    \caption{The \textit{p} orbital-decomposed MAE (in unit of meV) of Te$_{1/2}$ atom in \ce{Cr2Ge2Se3Te3} monolayer under $-6\%$ strain (a), $-4\%$ strain (b), $-2\%$ strain (c), 0\% strain (d), 2\% strain (e), and 4\% strain (f).}
    \label{fig:APL-Figure4}
\end{figure}

Furthermore, since the \ce{Te1} and \ce{Te2} atoms are equivalent, we denote them as Te$_{1/2}$. Fig. \ref{fig:APL-Figure4} shows the \textit{p} orbital-decomposed MAE of Te$_{1/2}$ atom under different strains. It can be seen that the contribution of Te$_{1/2}$ to MAE (MAE$_{\rm{Te}_{1/2}}$) is mainly influenced  by  $p_z / p_y$ and $p_x / p_y$  matrices, and it reaches its maximum (MAE$_{\rm{Te}_{1/2}}$ = 0.30 $-$ 0.09 = 0.21 meV) 
when the strain is equal to $-2\%$. This is consistent with the blue curve in Fig. \ref{fig:APL-Figure3}(b).

\begin{figure}
    \centering
    \includegraphics[width=0.99\linewidth]{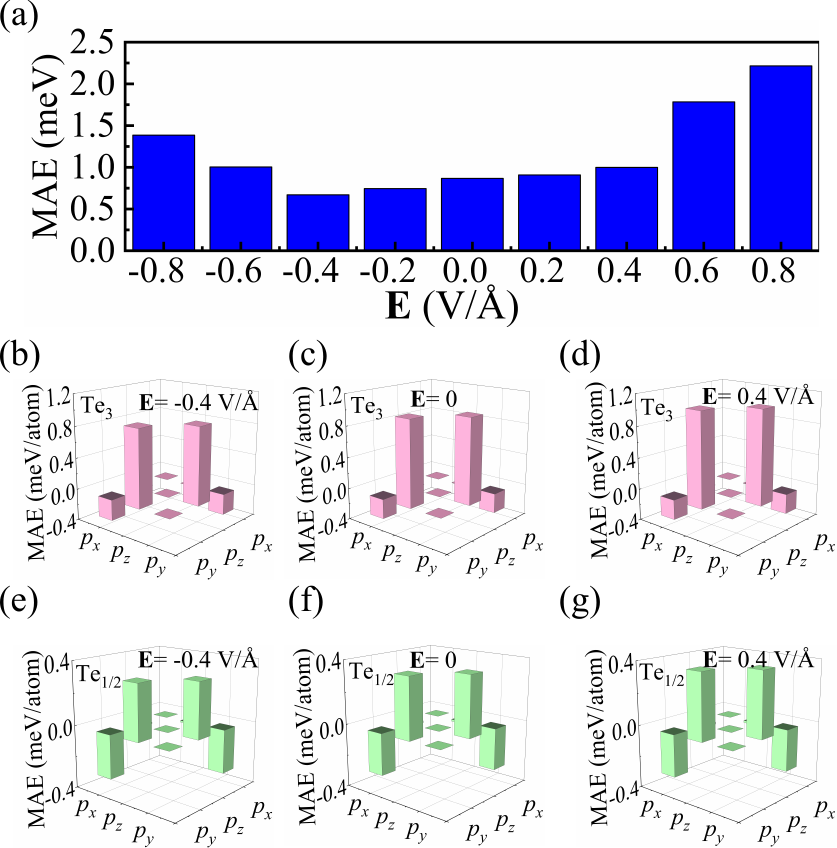}
    \caption{(a) The total MAE dependency on the electric field. The \textit{p} orbital-decomposed MAE of \ce{Te3} and Te$_{1/2}$ atoms in \ce{Cr2Ge2Se3Te3} monolayer under different field: $-0.4$ V/\AA\ in (b) and (e), 0 V/\AA\ in (c) and (f), and 0.4 V/\AA\ in (d) and (g).}
    \label{fig:APL-Figure5}
\end{figure}

Finally, we focus on investigating the regulation of MAE by electric field \textbf{E}. Here, a positive electric field is defined to be pointed away from the surface of the Janus ferromagnetic monolayer, which is along the [001] direction. Fig. \ref{fig:APL-Figure5}(a) illustrates the dependence of the MAE on the electric field. It is clear that the MAE is highly sensitive to the direction of the electric field, with positive and negative fields resulting in different MAE values. This confirms the non-centrosymmetry of the ferromagnet. Particularly in the range of strain from $-4\%$ to 4\%, the total MAE of the system increases monotonically, which is consistent with the increasing contribution of $p_z / p_y$ hybridization in the orbital-decomposed MAE of \ce{Te3} and Te$_{1/2}$ in Figs. \ref{fig:APL-Figure5}(b)-\ref{fig:APL-Figure5}(g). It is worth noting that the contribution of Te$_{1/2}$ atoms is significantly smaller than that of the \ce{Te3} atoms, further supporting the idea that the \ce{Te3} atom play a major role in MAE.


In summary, by combining the first-principles calculations with Monte Carlo simulations, we predict a Janus ferromagnet \ce{Cr2Ge2Se3Te3} monolayer in which reduced dimensionality, inversion symmetry breaking and exotic electronic and magnetic behaviors appear together. 
Strain is found to induce a magnetic transition between FM and AFM states, as well as a switch of the magnetization easy axis by 90$^{\circ}$. It is shown that the main contributor to MAE is the \ce{Te3} atoms, in which the competition between the two matrices $p_z/p_y$ and $p_x/p_y$ matrices plays a crucial role. Recent experiments have demonstrated the synthesis of Janus monolayers such as MoSSe~\cite{Jang2022-NPG-Asia-MoSSe} and SPtSe~\cite{Sant2020-npj-SPtSe}, suggesting that the synthesis of 2D \ce{Cr2Ge2Se3Te3} is likely to be achieved by molecular beam epitaxy (MBE) or liquid-phase cation exchange method in the near future. Our findings present a promising candidate for experimental research on 2D Janus magnets, and suggest that strain and electric fields are effective methods for tuning the electronic and magnetic properties for spintronics applications.

\textbf{Supplementary material. } See the supplementary material for the Wyckoff positions of the \ce{Cr2Ge2Se3Te3} monolayer (Tab. S1); The \textit{P}31\textit{m} and \textit{P}6\textit{mm} atomic structures of the \ce{Cr2Ge2Se3Te3} monolayer (Fig. S1); $A$, and $D_z$ as functions of strain for \ce{Cr2Ge2Se3Te3} unit cell (Fig. S2); Different magnetic orderings of the 2 $\times$ 2 $\times$ 1 cell for \ce{Cr2Ge2Se3Te3} (Fig. S3) and its $J_1$, $J_2$, $A$, $D_z$, and transition temperature as functions of strain (Fig. S4); Band gap of \ce{Cr2Ge2Se3Te3} dependence on ($U$, $J$) in the strain-free case, and the calculated critical strain for the metal-semiconductor transition dependence on ($U$, $J$) (Fig. S5).

\textbf{Acknowledgment. } This study is supported by Natural Science Basic Research Program of Shaanxi (Program No. 2024JC-YBQN-0038) and Extraordinary Grant of CSIC (No. 2025ICT122).

\textbf{Competing interests. } The authors declare no competing interests.

\textbf{Author Contributions. } Rui-Qi Wang: Computation, Writing - original draft, Analysis, Funding acquisition. Tengfei Cao: Editing. Tian-Min Lei: Resources, Supervision. Xie Zhang: Review, Editing. Yue-Wen Fang: Analysis, Writing - review \& editing, Supervision, Funding acquisition.

\textbf{Data availability.} All data supporting the findings of this study are available within the article, as well as the supplementary information.





\clearpage\newpage
\bibliographystyle{apsrev4-2}
%

\clearpage
\appendix
\onecolumngrid  

\section*{Supporting Information}
\addcontentsline{toc}{section}{Supporting Information} 

\setcounter{figure}{0} 
\renewcommand\thefigure{S\arabic{figure}} 
\renewcommand{\theHfigure}{S\arabic{figure}}

\setcounter{table}{0} 
\renewcommand\thetable{S\arabic{table}} 

Table \ref{Tab:SI1} shows the lattice parameters and Wyckoff positions of the \ce{Cr2Ge2Se3Te3} monolayer.

\begin{table}[htb]
    \caption{The Wyckoff positions of the \ce{Cr2Ge2Se3Te3} monolayer. The lattice constant is $a$ = 6.634 \AA, $b$ = 6.634 \AA, $c$ = 30 \AA, $\alpha$ = $\beta$ = 90$^\circ$, and $\gamma = 120^\circ$.}
    \label{Tab:SI1}
    \centering
    \begin{tabular}{ccccccccc} 
        \hline 
        Atoms& Wyckoff& x& y&& z&\\
        \hline
        Te& 3c&0.6147950000&0.0000000000& &0.5887043333& \\
        Se& 3c&0.3560980000&0.0000000000& &0.4743100000& \\
        Cr& 2b& 0.3333333333&0.6666666667& &0.5255505000& \\
        Ge1& 1a&0.0000000000&0.0000000000& &0.5729650000& \\
        Ge2& 1a&0.0000000000&0.0000000000& &0.4928540000& \\
        \hline
    \end{tabular}
\end{table}

\begin{figure}[htp]
    \centering
    \includegraphics[width=0.6\textwidth]{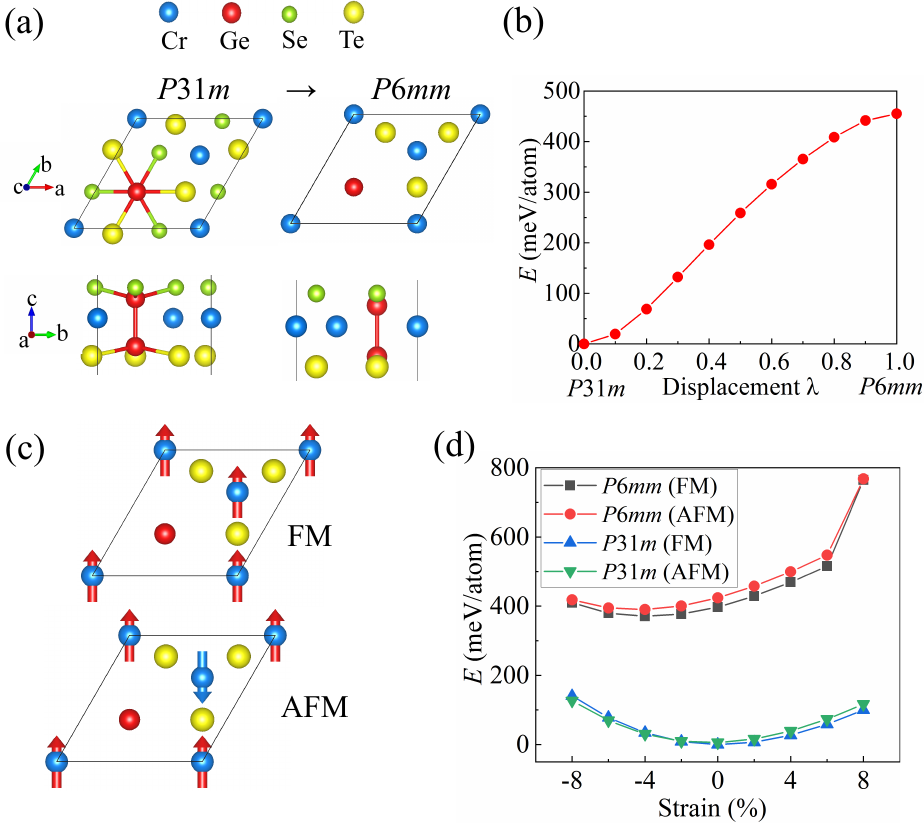}
    \caption{(a) \textit{P}31\textit{m} and \textit{P}6\textit{mm} atomic structures of the \ce{Cr2Ge2Se3Te3} monolayer. The Cr, Ge, Se, and Te atoms are represented by blue, red, green, and yellow spheres, respectively. (b) The energy transition from the low-symmetry \textit{P}31\textit{m} structure to the high-symmetry \textit{P}6\textit{mm} structure. (c) FM and AFM orderings of the \textit{P}6\textit{mm} structure of the \ce{Cr2Ge2Se3Te3} monolayer. (d) The system energies as a function of the strain.}
    \label{APL-figs1}
\end{figure}

Fig. \ref{APL-figs1}(a) shows the $P6mm$ phase of \ce{Cr2Ge2Se3Te3}. In the calculations of  $P6mm$ phase, the considered magnetic structures are shown in Fig. \ref{APL-figs1}(c).
Fig. \ref{APL-figs1}(b) shows the energy variations versus the normalized displacements between the two structures, demonstrating a continuous and smooth energy curve.
In addition, \textit{P}31\textit{m} phase is always lower than $P6mm$ phase in energy regardless of the application of strain.

\begin{figure}[htp]
    \centering
    \includegraphics[width=0.4\textwidth]{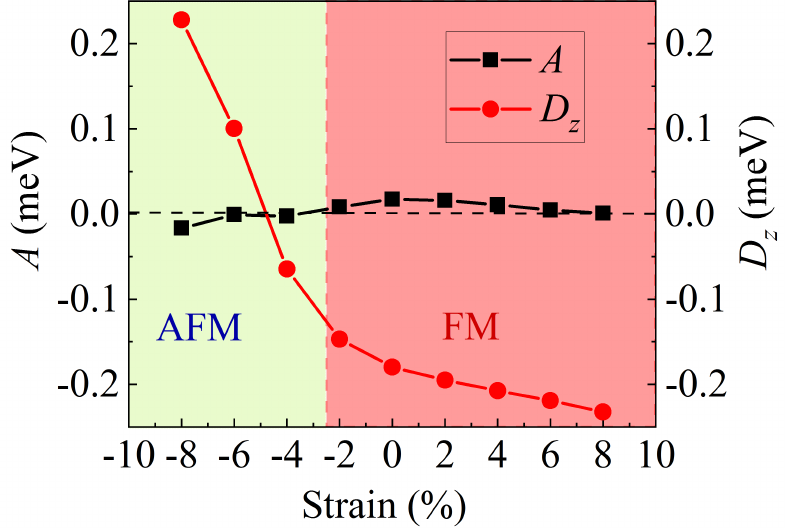}
    \caption{The single-ion anisotropy parameter \( A\)  and DMI strength  \( D_z\) as functions of strain.}
    \label{APL-figs2}
\end{figure}

Figs. \ref{APL-figs2} shows the single-ion anisotropy (SIA) parameter \( A\) and Dzyaloshinskii-Moriya 
interaction (DMI) strength \( D_z\) as functions of strain. The single-ion magnetic anisotropy parameter \( A\)  is calculated by setting the following four spin configurations\cite{Xianghongjun-2013}: \((i) S_1=(0, 0, S)  , (ii) S_1=(0, 0, -S), (iii) S_1=(S, 0, 0), (iv) S_1=(-S, 0, 0).\) Since there are two Cr atoms, the other spin site of Cr is set to  \( S_2=(0, S, 0) \) and keep the same for the four spin states. Then,  
\begin{equation}
A=\frac{ E_1+E_2-(E_3+E_4)} {2S^2}
\end{equation}

For DMI strength,  we calculate the out-of-plane component \( D_z\)  between the spin site 1 and spin site 2 ( \( S_1\) and \( S_2\) ) by setting the following four spin configurations\cite{Xianghongjun-2013}: \((i) S_1=(S, 0, 0), S_2=(0, S, 0)  , (ii) S_1=(S, 0, 0), S_2=(0, -S, 0), (iii) S_1=(-S, 0, 0), S_2=(0, S, 0), (iv) S_1=(-S, 0, 0), S_2=(0, -S, 0).\) We note that the spins of all the other sites are the same and along the z direction. The spin interaction energies of the four spin configurations are denoted by \( E_1\) , \( E_2\) , \( E_3\)  , and \( E_4\) .  \( D_z\) can be calculated by:
\begin{equation}
D_z=\frac{ E_1+E_4-(E_2+E_3)} {4S^2}
\end{equation}

We also utilize the four states method\cite{Xianghongjun-2013} to calculate \( D_x\) and \( D_y\), and find that their quantities are nearly zero. 

What should be noted is that when multiplied by \(2S^2\) and the number of magnetic
atoms, the SIA parameter \( A\) is also an order of magnitude smaller than MAE. This is due to the fact that the MAE is dominated by collective effects, especially in magnets with itinerant electrons. However, the SIA model is based on the local magnetic moment assumption. 

\begin{figure}[htp]
    \centering
    \includegraphics[width=0.68\textwidth]{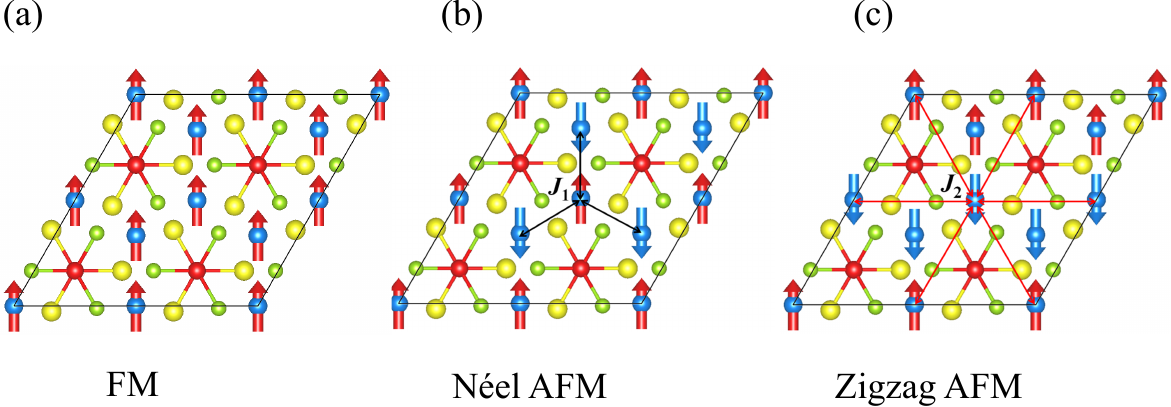}
    \caption{Different magnetic orderings of the 2 $\times$ 2 $\times$ 1 cell of ferromagnetic \ce{Cr2Ge2Se3Te3}: (a) FM, (b) \Neel\ AFM, and (c) Zigzag AFM. $J_1$ and $J_2$ represent the nearest and next-nearest neighbor exchange couplings.}
    \label{APL-figs3}
\end{figure}

Figs. \ref{APL-figs3} shows the FM ordering, the \Neel\ AFM ordering, and the Zigzag AFM ordering in a 2 $\times$ 2 $\times$ 1 cell. When the nearest-neighbor exchange interaction and the next nearest neighbor interaction are both considered, the total energies of \ce{Cr2Ge2Se3Te3} with FM, \Neel\ AFM, and Zigzag AFM orderings can be expressed as:
\begin{equation}
H_{F M}=-12 J_1 S^2-24 J_2 S^2-A S^2+E_{other}
\end{equation}
\begin{equation}
H_{N\acute{e}el}=12 J_1 S^2-24 J_2 S^2-A S^2+E_{other}
\end{equation}
\begin{equation}
H_{Zigzag}=-4 J_1 S^2+8 J_2 S^2-A S^2+E_{other}
\end{equation}
Where \textit{H}$_{FM}$, \textit{H}$_{N\acute{e}el}$ and \textit{H}$_{Zigzag}$ are the total energy of FM, \Neel\ AFM, and Zigzag AFM orderings, respectively. $J_1$ is the nearest-neighbor exchange parameter, $J_2$ is the next nearest-neighbor exchange parameter, $A$ represents the single-ion anisotropy, and \textit{E}$_{other}$ includes the DMI strength $D_z$ and the total energy without magnetic coupling. 

\begin{figure}[htp
]
    \centering
    \includegraphics[width=0.9\textwidth]{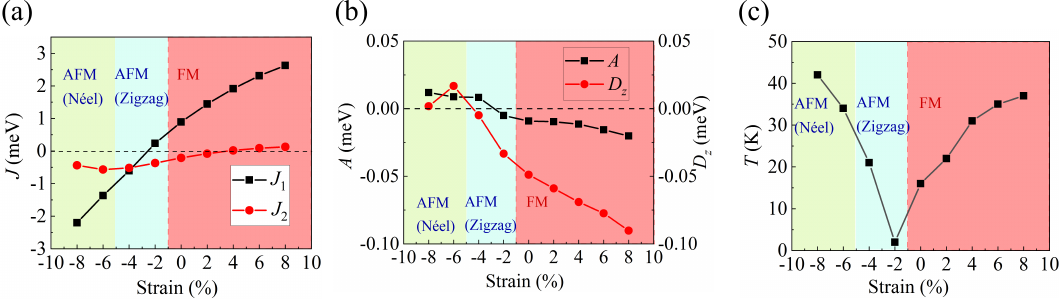}
    \caption{(a) The nearest-neighbor exchange parameter $J_1$ and the next nearest-neighbor exchange parameter $J_2$ as functions of the strain. (b) The single-ion anisotropy parameter \( A\)  and DMI strength  \( D_z\) as functions of strain. (c) The transition temperature as a function of the strain.}
    \label{APL-figs4}
\end{figure}

Figs. \ref{APL-figs4}(a)-\ref{APL-figs4}(b) show the $J_1$, $J_2$, $A$ and $D_z$ values. $J_1$ and $J_2$ are obtained by solving the equations, while $A$ and $D_z$ are obtained using
the four-state energy-mapping analysis\cite{Xianghongjun-2013}. Here we assume that all spins of Cr atoms are equivalent. Through Monte Carlo simulations, the transition temperature dependencies on strain are plotted in Fig. \ref{APL-figs4}(c). It can be seen from Fig. \ref{APL-figs4}(a) that with the increase of strain, $J_1$ increases rapidly and $J_2$ changes limitedly, reflecting the limited influence of the next-nearest-neighbor exchange interaction. Fig. \ref{APL-figs4}(b) reveals that $A$ and $D_z$ are both an order of magnitude smaller than $J_1$ ($J_2$). The transition temperature $T$ does not exceed 42 K in Fig. \ref{APL-figs4}(c), which is not much different from the $T$ when only the nearest neighbor is considered.

\begin{figure}[htp]
    \centering
    \includegraphics[width=0.7\textwidth]{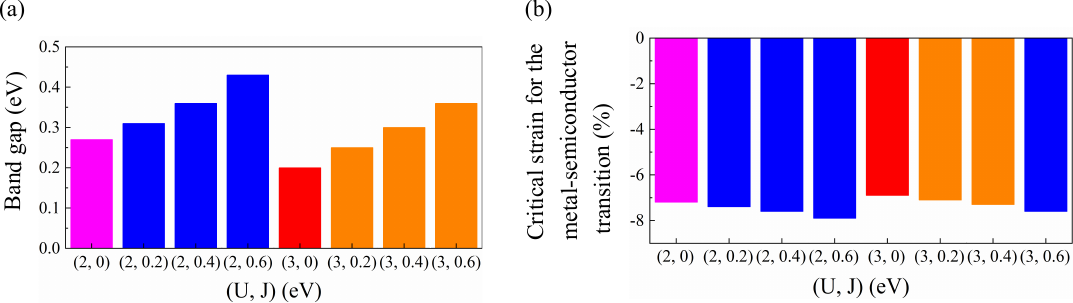}
    \caption{(a) Band gap of \ce{Cr2Ge2Se3Te3} dependence on ($U$, $J$) with 0\% strain; (b) Critical strain for the metal-semiconductor transition dependence on ($U$, $J$). In these LDA+$U$ calculations, Liechtenstein et al~\cite{Liechtenstein1995} approach was used.}
    \label{FIG-response1}
\end{figure}

Figs. \ref{FIG-response1}(a) and \ref{FIG-response1}(b) show the calculated band gap and critical strain for the metal-semiconductor transition dependence on ($U$, $J$), respectively. It can be seen that when the $U$ value is constant, the larger the $J$ value, the larger the band gap, and therefore the greater the strength of compressive strain required to close it. In all calculations, the band gap is always around 0.3 eV in the strain-free state, and the critical strain for the metal-semiconductor transition is persistently between -7\% and -8\%.

\end{document}